\documentclass[prd,twocolumn,amsmath]{revtex4}
\usepackage{amssymb,amsmath}
\usepackage{color}
\usepackage{graphicx}

\newcommand{\rmax}{\rho_{max}}
\usepackage[normalem]{ulem}

\usepackage[dvipsnames]{xcolor}

\begin{document}
\title{Quantum Fields, Geometric Fluctuations, and the Structure of Spacetime}

\author{S. Carlip}
\email[]{carlip@physics.ucdavis.edu}
\affiliation{Department of Physics, University of California, Davis, CA 95616, USA}

\author{R. A. Mosna}
\email[]{mosna@unicamp.br}
\affiliation{Departamento de Matem\'atica Aplicada, Universidade Estadual de Campinas, 
13083-859, Campinas, S\~ao Paulo, Brazil}

\author{J. P. M. Pitelli}
\email[]{pitelli@unicamp.br}
\affiliation{Departamento de Matem\'atica Aplicada, Universidade Estadual de Campinas, 
13083-859, Campinas, S\~ao Paulo, Brazil}

\begin{abstract}

Quantum fluctuations of the vacuum stress-energy tensor are highly non-Gaussian, and can have unexpectedly large effects on spacetime geometry.  In this paper, we study a two-dimensional dilaton gravity model coupled to a conformal field, in which the distribution of vacuum fluctuations is well understood. In this model, the fluctuations  of the matter field are responsible for the fluctuations  of the geometry itself. By analyzing the geodesic deviation in this model, we show that a pencil of massive particles propagating on this fuzzy spacetime eventually converges and collapses.  This is consistent with our earlier analysis of null geodesics in [Phys.~Rev.~Lett.\ 107, 021303 (2011)].

\end{abstract}

\maketitle

\section {Introduction}

Quantum fluctuations of the vacuum energy, when connected to geometry via the Einstein field equations, cause fluctuations of the spacetime itself. These have the potential to disturb the motion of particles and even the underlying causal structure. In Ref.~\cite{carlip1}, we studied the impact of the vacuum fluctuations of a conformal field on the causal structure of spacetime. To do this, we analyzed the Raychaudhuri equation for a pencil of light in a 
two-dimensional dilaton gravity model for which the probability distribution for the fluctuations is exactly known; the dilaton field in this case played the role of a transverse area in the ``missing'' dimensions. In this context, we showed that the fluctuations of the stress-energy tensor lead to a sharp focusing of light cones near the Planck scale, breaking up the causal structure of spacetime  at such small scales. 

Additional evidence for this phenomenon coming from perturbative algebraic quantum field theory was obtained in Ref.~\cite{Drago}.  The connection between vacuum fluctuations  and spacetime geometry was further studied in Ref.~\cite{vieira}. Since the exact probability distribution  for vacuum fluctuations is  not known in four dimensions, only the variance of the relative velocity and the mean  squared distance fluctuation could be obtained, through Riemann tensor correlation functions, in this case.  In two spacetime dimensions, on the other hand, the exact probability distribution for fluctuations of the stress-energy tensor is known, at least for conformal fields in Minkowski spacetime~\cite{fewster,fewster2,fewster3}. Finite results require that the stress-energy operator be smeared by a test function, but for a wide variety of smearings in time, the probability distribution for the fluctuations is given by a shifted gamma distribution~\cite{fewster2}. The smearing introduces an arbitrary time scale, with quantum gravity effects arising as one approaches the Planck scale.

In this paper, we explore another aspect of  vacuum fluctuations in two spacetime dimensions, which was neglected in~\cite{carlip1}:  besides the direct effect on the dilaton, the vacuum fluctuations of the stress-energy tensor couple to the metric, inducing fluctuations of the spacetime itself.  Pure Einstein gravity has no dynamics in two dimensions, since the Einstein-Hilbert action is a topological invariant. In dilaton gravity, though, the spacetime curvature is determined by the dilaton potential. Thus, once we find how the dilaton responds to vacuum fluctuations of the quantum field, we can determine the curvature fluctuations and their effect on particle trajectories.  This is the main purpose of this work. 

\section {The model}

Two-dimensional dilaton gravity can be obtained by  dimensional reduction 
from higher-dimensional general relativity.  Under such a reduction, the dilaton $\varphi$ 
is essentially the transverse area element.   In a previous paper~\cite{carlip1}, we 
considered the direct effect of vacuum fluctuations on $\varphi$, viewed as an area, by 
using a version of the Raychaudhuri equation in which the fluctuations acted as a 
stochastic noise term.  Here we consider a process of a more geometric nature: fluctuations of 
the vacuum stress-energy tensor induce fluctuations of the curvature, which in turn
affect the behavior of timelike geodesics.  As we shall see, the two analyses lead to a 
consistent picture, in which vacuum fluctuations at the Planck scale lead the ``collapse'' of 
a pencil of geodesics at that scale.

In a dilatonic theory, with appropriate redefinitions, it is always possible to bring the 
action into the form~\cite{gegenberg,navarro}
\begin{equation}
S=S_V+S_M,
\label{main action}
\end{equation}
with
\begin{equation}
S_V=\int{d^2x\sqrt{-g}\left[\varphi R+V(\varphi)\right]}
\end{equation}
being the geometrical action and 
\begin{equation}
S_M=\int{d^2x\sqrt{-g}\mathcal{L}_M}
\end{equation}
being the action for the matter fields. In what follows, we will take $S_M$ to describe
a conformal field with central charge $\bar{c}=1$.  As we will 
see later, given a characteristic scale $\bar{\tau}$ (the smearing scale for the quantum field), the Ricci scalar will be small compared to $1/\bar{\tau}^2$, and the random geometry will be close to that  of flat spacetime. We will therefore approximate the  vacuum fluctuations of the matter stress-energy tensor by their flat spacetime distribution.

The right-moving/left-moving components of the smeared 
stress-energy tensor have  probability distributions for individual measurements given 
by a shifted gamma distribution~\cite{fewster3},
\begin{equation}
P(T_{R/L}=\omega)=\Theta(\omega+\omega_0)
\frac{\beta^{\alpha}(\omega+\omega_0)^{\alpha-1}}{\Gamma(\alpha)}e^{-\beta(\omega+\omega_0)},
\label{probdist}
\end{equation} 
with
\begin{equation}
\omega_0=\frac{1}{48\pi\sigma^2},\,\,\,\alpha=\frac{1}{24},\,\,\,\beta=2\pi \sigma^2,
\label{parameters}
\end{equation}
where  $\sigma^2=(\Delta t)^2+(\Delta x)^2$, with $\Delta t$ and $\Delta x$ being the characteristic widths of the smearing in time  and space.  For the purpose of numerical simulation, we divide the spacetime in our model into rectangular patches such that $\Delta t=\Delta x=\bar{\tau}$, where $\bar{\tau}$ is an arbitrary scale, which may be identified later with the Planck scale $\tau_p$ if we want to investigate quantum gravity effects. We then take the fluctuations to act independently on each of these patches.  This is not quite correct---fluctuations in nearby patches are correlated---but as shown in~\cite{fewster}, these correlations fall off very rapidly with distance.

The connection between geometry and vacuum fluctuations comes into being as follows. Varying the action~(\ref{main action}) with respect to $\varphi$ leads to
\begin{equation}
R=-V'(\varphi).
\label{curvature phi}
\end{equation}
This equation determines the curvature of the two-dimensional spacetime in terms of the dilaton
 $\varphi$. The equation for the dilaton field is obtained by varying the action 
 with respect to the metric $g_{\mu\nu}$, which leads to
\begin{equation}
\nabla_{\mu}\nabla_{\nu}\varphi=\frac{1}{2}g_{\mu\nu}V(\varphi)+g_{\mu\nu}T-T_{\mu\nu},
\label{main}
\end{equation} 
where $T_{\mu\nu}$ is the stress-energy tensor associated with the matter field, and $T=0$  since we are dealing with a conformal field. Here we consider a $V(\varphi)$ to be small near $\varphi=0$, so  
\begin{equation}
V(\varphi)=\frac{1}{2}V_0\varphi^2.
\end{equation}

Let us write the spacetime metric in terms of a conformal factor
\begin{equation}
\label{metric}
ds^2=e^{\rho(t,x)}(dt^2-dx^2) .
\end{equation}
In two dimensions, such form can always be achieved, at least locally, through a choice of coordinates.
The equations of motion (\ref{curvature phi}) and~(\ref{main}) then yield a complicated set of PDEs to be solved for 
$\rho(t,x)$.  To first order in $\rho(t,x)$, however, they simplify to
\begin{equation}
\begin{aligned}
&\frac{\partial^4\rho}{\partial \eta^3\partial\xi}=-\frac{V_0}{4}T_R,\\
&\frac{\partial^4\rho}{\partial \eta^2\partial\xi^2}=0,\\
&\frac{\partial^4\rho}{\partial \eta\partial\xi^3}=-\frac{V_0}{4}T_L,
\end{aligned}
\end{equation}
where $\eta =t-x$ and $\xi = t+x$. The general solution to this system of equations is given by
\begin{equation}
\begin{aligned}
\rho(\eta,\xi)=&-\frac{V_0\eta\xi}{24}\left(T_R\eta^2+T_L\xi^2\right)+\eta\xi(c_1+c_2\eta+c_3\xi)\\
&+f_1(\eta)+f_2(\xi).
\end{aligned}
\end{equation}

The solution that reduces to the Minkowski metric (in the usual coordinates) when $T_R=T_L= 0$ is thus 
\begin{equation}
\rho(\eta,\xi)=-\frac{V_0\eta\xi}{24}\left(T_R\eta^2+T_L\xi^2\right),
\label{rho}
\end{equation}
which in turn gives rise to a scalar curvature of the form
\begin{equation}
R(\eta,\xi)=\frac{V_0}{2}\left(T_R\eta^2+T_L\xi^2\right)e^{\frac{V_0}{24}\eta\xi\left(T_R\eta^2+T_L\xi^2\right)}.
\label{R}
\end{equation}
These expressions are valid on each rectangular patch of dimensions $\Delta x=\Delta t=\bar{\tau}$, in the center of which we momentarily put the origin of the coordinates.

\begin{figure}[h]
\centering
\includegraphics[width=0.49\columnwidth]{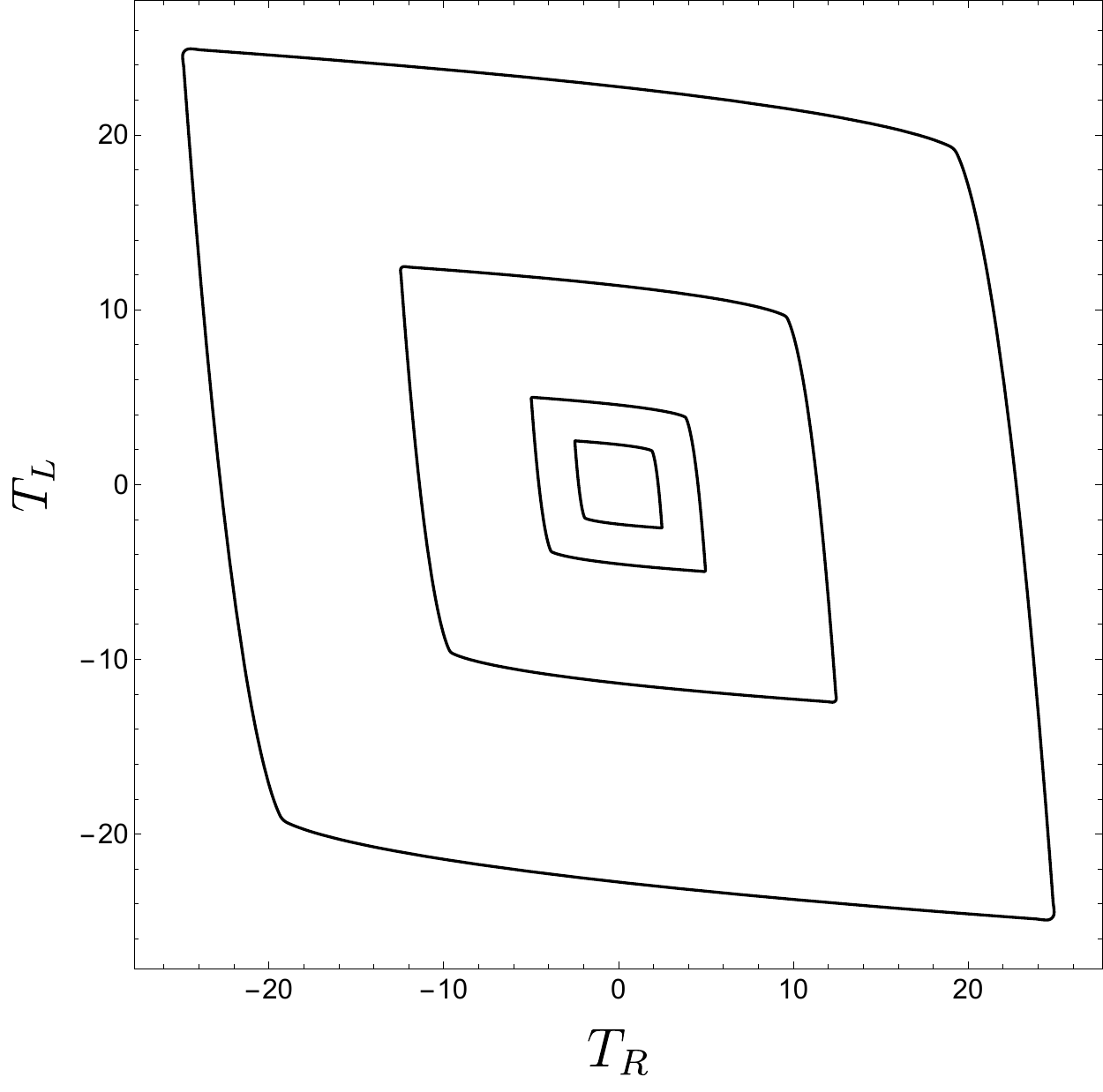}
\includegraphics[width=0.49\columnwidth]{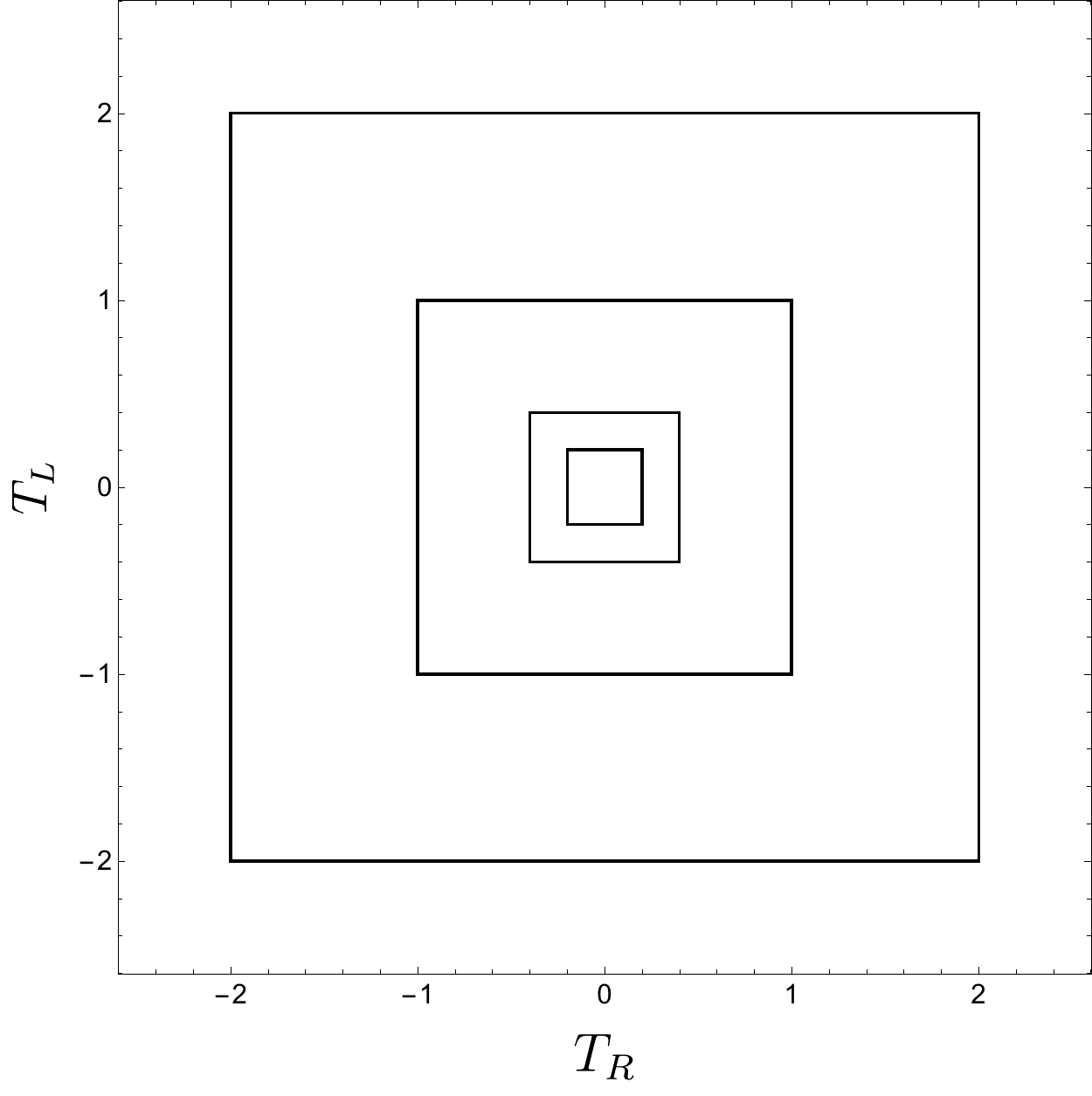}
\includegraphics[width=0.49\columnwidth]{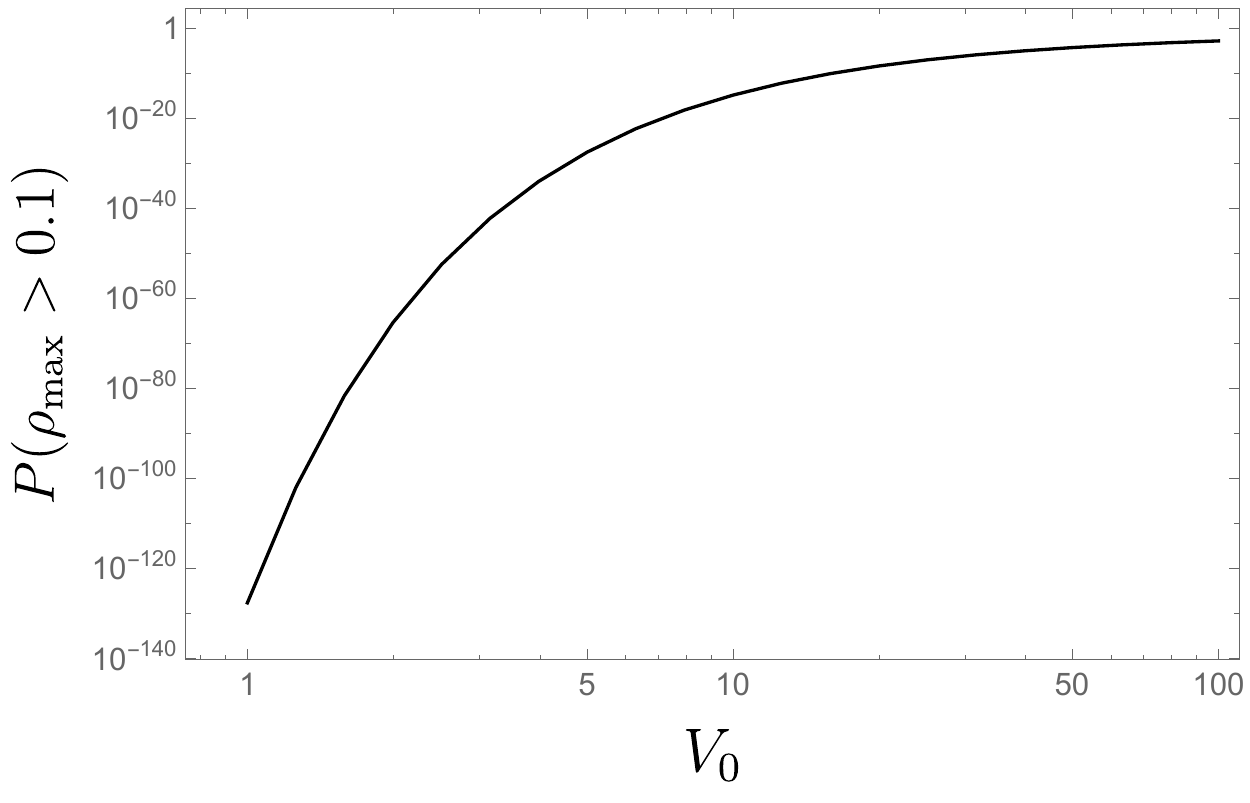}
\includegraphics[width=0.49\columnwidth]{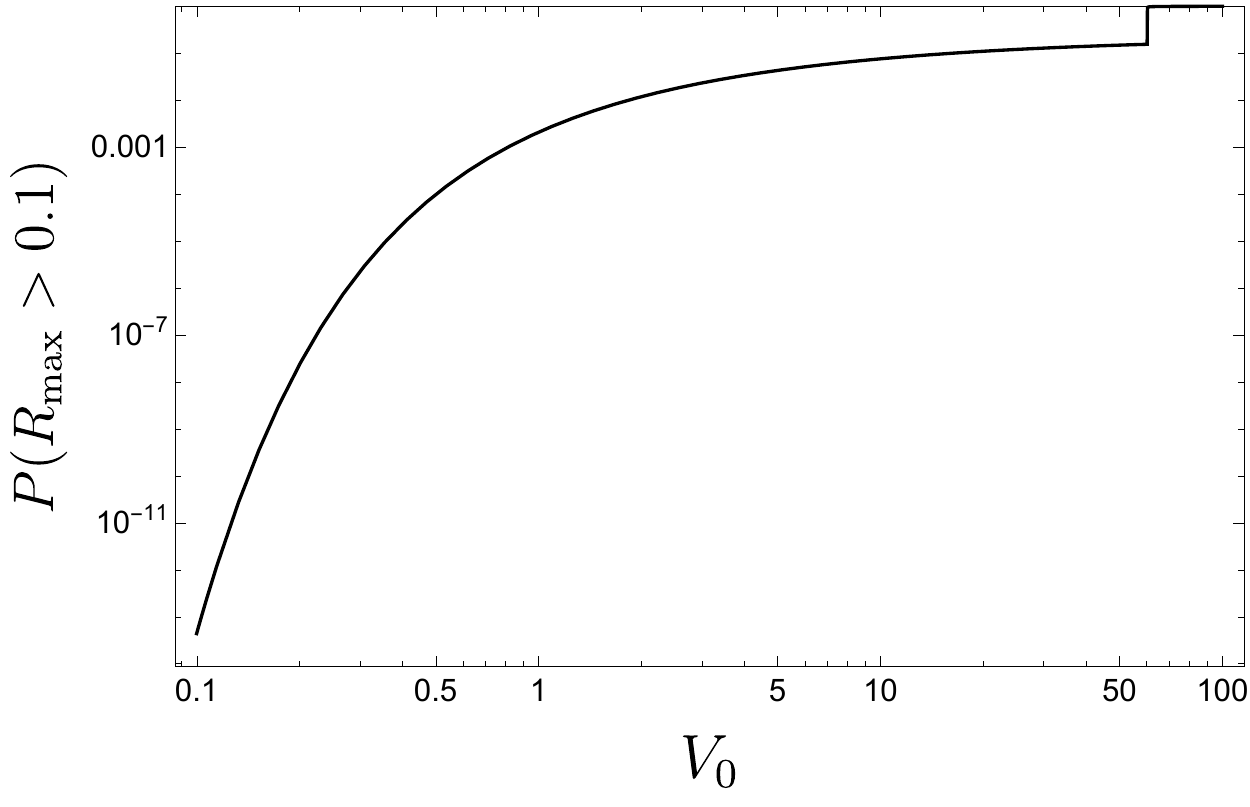}
\caption{Top left: curves in the $T_R T_L$ plane for which $\rho_{max}=0.1$ for $V_0=1$ (outermost curve), $V_0=2$, $V_0=5$, $V_0=10$, $V_0=50$ and $V_0=100$ (innermost curve).
Top right: the same for the condition $R_{max}=0.1$. The region for which $\rho_{max}>0.1$, resp. $R_{max}=0.1$, lies outside the corresponding curves. Bottom: probability of $\rmax>0.1$ (left) and $R_{max}>0.1$ (right) as a function of $V_0$. All plots are in units $\bar{\tau}=1$.}
\label{contornos}
\end{figure}

We want to work in a regime in which $\rho\ll 1$ and the spacetime is nearly flat. In each patch, the absolute values of $\rho$ and $R$ in Eqs.~(\ref{rho}) and (\ref{R}) assume maximum values, $\rho_{max}$ and $R_{max}$, at some point at the boundary of the patch. Fig.~\ref{contornos} shows the curves in the $T_R T_L$-plane that correspond to the condition $\rho_{max}=0.1$ and $R_{max}=0.1$ for various choices of $V_0$.

Although $\rho$ and $R$ can always assume arbitrarily large values, we see from Fig.~\ref{contornos} that the regions in the $T_R T_L$ plane corresponding to $\rho_{max}>0.1$ or $R_{max}>0.1$ get smaller as $V_0$ decreases. Using Eq.~(\ref{probdist}) to define a joint probability function $P(T_R,T_L)=P(T_R)P(T_L)$, we find, for $V_0=1$, that $P\left(\rho_{max}>0.1\right)\sim10^{-128}$ and $P\left(R_{max}>0.1\right)\sim2\times10^{-3}$. Both of these values can be made arbitrarily small by choosing a small enough $V_0$~\cite{endnote1}.

\section{Collapse time}

We now analyze the behavior of a congruence of timelike particles in this spacetime. We assume this congruence is initially given by a pencil of particles with velocity $v=0$ with respect to the laboratory (the frame in which the stress-energy tensor is measured). The particles then follow geodesics determined by the metric given by Eqs.~(\ref{metric}) and (\ref{rho}). {As such, they deviate from each other in response to the scalar curvature in Eq.~(\ref{R}).} Alternatively, we can solve the Raychaudhuri equation for this congruence,
\begin{equation}
\frac{d\theta(\tau)}{d\tau}=-\theta(\tau)^2-\frac{1}{2} R(x(\tau)),
\label{eq:ray}
\end{equation} 
where $\tau$ is the proper time. In terms of the coordinate time $t$ of the laboratory frame, this becomes
\begin{equation}
\label{eq:theta}
\frac{d\theta(t)}{dt}=-e^{\rho(t,x(t))} \sqrt{1-\dot{x}(t)^2} 
\left[\theta(t)^2+\frac{1}{2} R(t,x(t)) \right],
\end{equation}
where $x(t)$ is determined by the coordinate expression of the geodesic equation with initial coordinate velocity $\dot{x}(0)=0$. As a result, we have a complicated system of two coupled nonlinear ODEs for $\theta(t)$ and $x(t)$, with parameters $T_R$ and $T_L$ that vary from patch to patch.

\subsection*{Numerical Solution}

We have solved this system of equations numerically in steps $\Delta t=\bar{\tau}$, with the initial condition at each step taken from the previous one and with $\theta(0)=0$, $x(0)=0$ and $\dot{x}(0)=0$, that is, an initially parallel pencil of particles initially at rest.  At each step, a numerical value for the vacuum fluctuation $T_{R/L}$ is randomly chosen using the probability  distribution (\ref{probdist}). We evolve the solution until $\theta$ diverges to $-\infty$. We call the time $t_c$ for which $\theta(t_c)\to-\infty$ the ``collapse time'' of the solution.

\begin{figure}[h]
\centering
\includegraphics[width=\columnwidth]{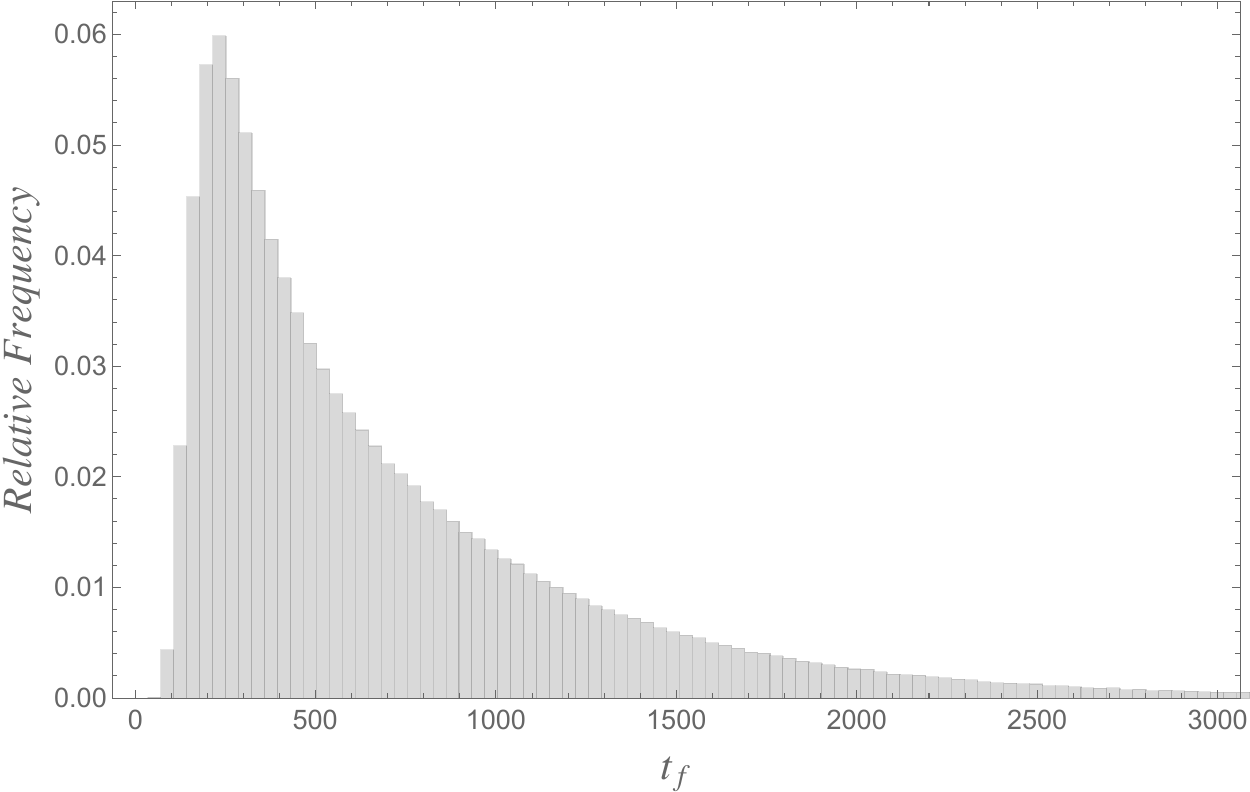}
\caption{Histogram of {collapse} times for $V_0=1$ in units of $\bar{\tau}$.}
\label{hist}
\end{figure}

\begin{figure}[h]
\centering
\includegraphics[width=0.75\columnwidth]{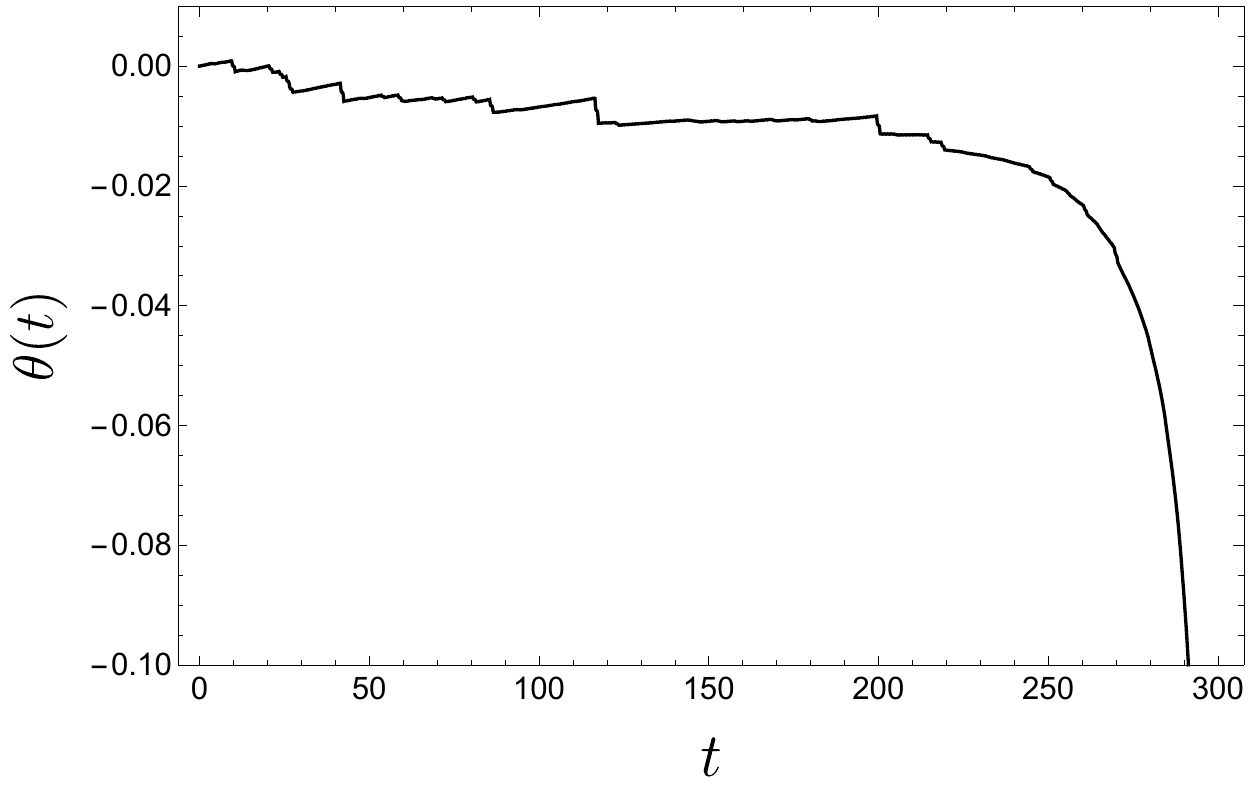}
\includegraphics[width=0.75\columnwidth]{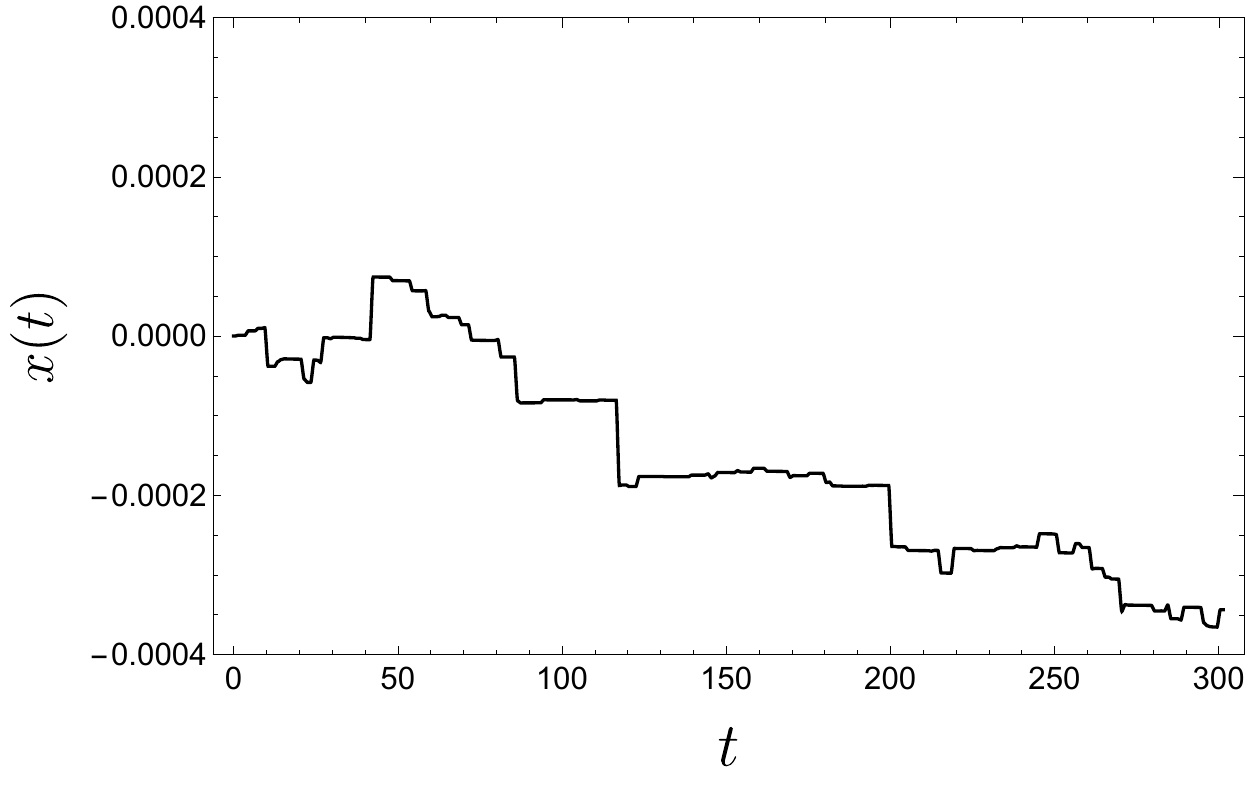}
\caption{Examples of trajectories for $\theta(t)$ (above) and $x(t)$ (below). The collapse time for this particular experiment was $\sim 300$, with $V_0=1$  in units of $\bar{\tau}$.}
\label{traj}
\end{figure}

The collapse time is itself a random variable, with a distribution we reconstruct by performing a large number $N$ of numerical experiments. The results for $N=10^6$ are shown in Fig.~\ref{hist} for $V_0=1$ (in units of $1/\bar{\tau}^2$). The associated mean collapse time is $t_c~\sim 718$, with a standard deviaton $611$ (in units of $\bar{\tau}$). Typical trajectories for $x(t)$ and $\theta(t)$ for this case are shown in Fig.~\ref{traj}. 

\begin{figure}[h]
\centering
\includegraphics[width=1\columnwidth]{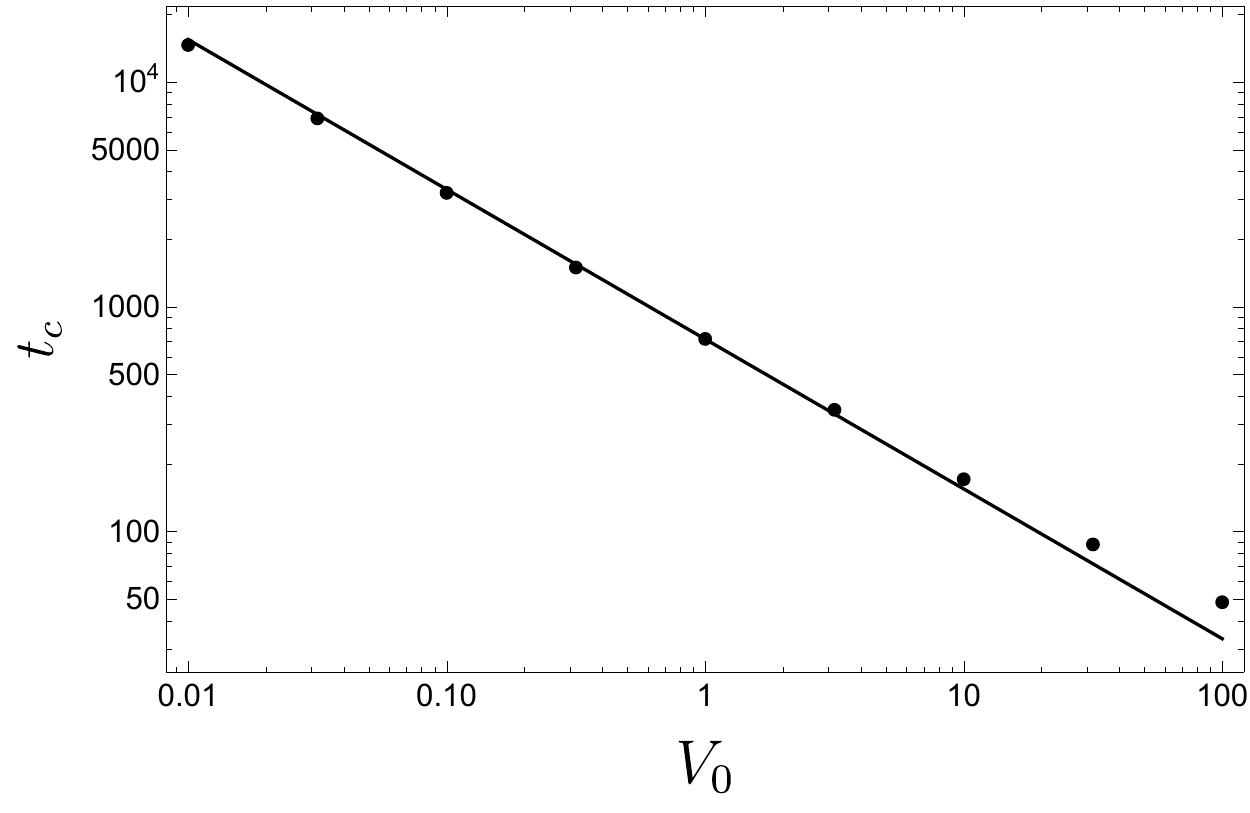}
\caption{Mean times of collapse as a function of $V_0$. The adjusted curve is given $t_c =718 \, V_0^{-2/3}$ (in units  
$\bar{\tau}=1$).}
\label{fit}
\end{figure}

It is worth emphasizing that although this value of $t_c$ is not particularly small in terms of $\bar\tau$, small times of collapse do play an important role in this model, as Fig.~\ref{hist} shows. In fact, the probability that $t_c<100\bar{\tau}$ is $\sim 0.25\%$ in this case, which is small but  non-negligible.  {One would find smaller values} for $t_c$ by considering larger values of $V_0$, as shown in Fig.~\ref{fit}. 

The problem with taking larger values for $V_0$ is the associated growth of $\rho_{max}$ and $R_{max}$. The condition $\rho\ll1$ is essential to our model, as discussed in the preceding section. As shown in Fig.~\ref{contornos}(c), though, this will hold even for $V_0$ as large as $\sim50$ (in units $\bar\tau=1$).   More importantly, the assumption of small $R$ was implicitly used when the flat spacetime distribution (\ref{probdist}) was adopted for the vacuum fluctuations $T_{R/L}$. {In contrast to $\rho_{max}$}, the value of $R_{max}$ is much more sensitive to the increase of $V_0$, as shown in Fig.~\ref{contornos}. 

\begin{figure}
\includegraphics[width=\columnwidth]{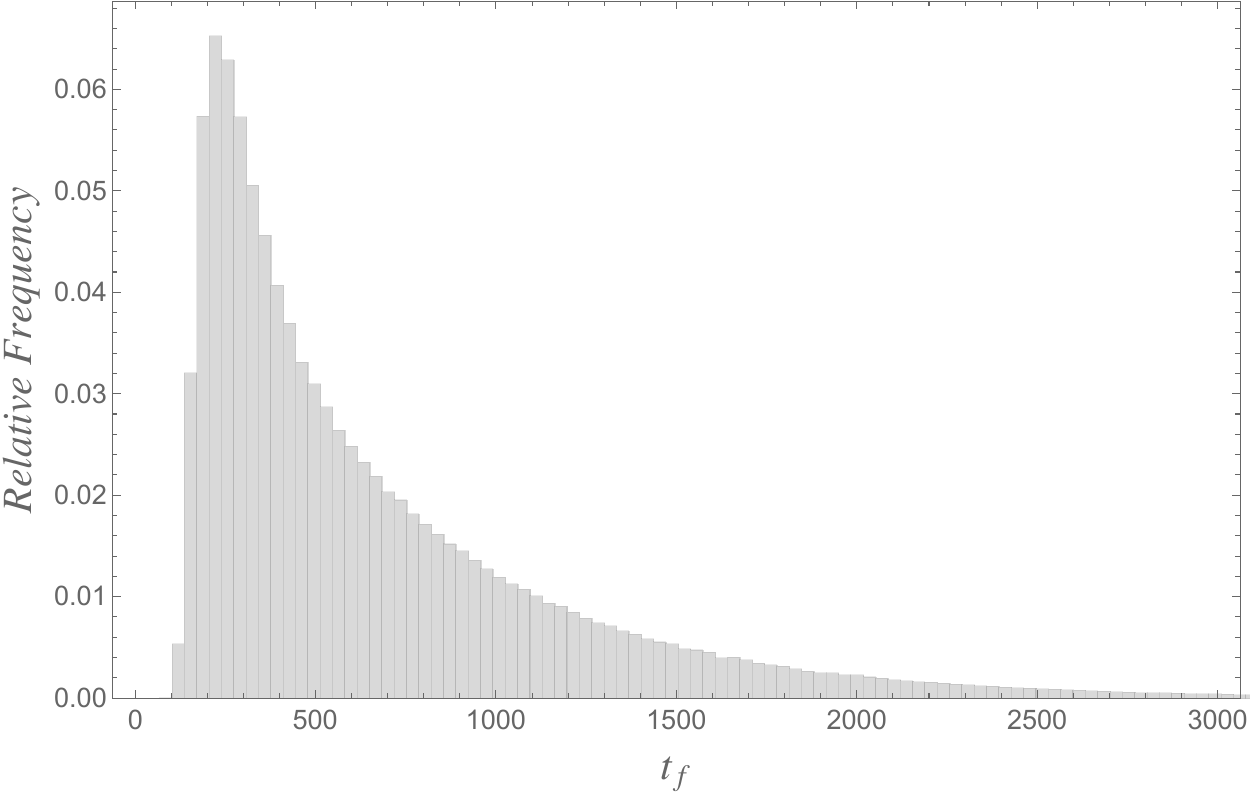}
\caption{Histogram for the collapse time using a discrete probability distribution for $T_{L/R}$ with only two possible outcomes, $\pm e_0$, each with probability $1/2$. The value of $e_0$ was chosen such that the mean and variance are the same those of Fig.~\ref{hist}. The number of experiments is again $10^6$. The mean time of collapse is $684$, with a standard deviation $565$ (in units $\bar{\tau}=1$). Note the similarity with the histogram of Fig.~\ref{hist}.}
\label{histcoin}
\end{figure}

Our results, however, are fairly independent of this detailed form of the probability distribution for $T_{R/L}$. In fact, any reasonable distribution with the same mean and standard deviation as the shifted gamma of Eq.~(\ref{probdist})
should lead to similar results.
To see this, suppose that the probability of drawing an energy more negative than a certain value, say $-\omega_0$, is essentially zero~\cite{endnote2}. This sets a lower negative bound for the scalar curvature, say $-R_0$, from Eq.~(\ref{R}). It then follows from Eq.~(\ref{eq:ray}) that $\theta$ will never grow bigger than $\theta_c=\sqrt{R_0}$ and that, once $\theta$ falls bellow $-R_0$, it will inevitably collapse toward $-\infty$.  To illustrate this point, we show in Fig.~\ref{histcoin} the histogram for the collapse times using a discrete distribution for $T_{L/R}$ with only two possible outcomes, $\pm e_0$, each with probability $1/2$ (where $e_0$ was chosen so that the mean and variance are the same as before). 
Our results thus essentially follow from the nonlinear nature of the Raychaudhuri equation; the exact details of the fluctuations are not crucial for our analysis.

Finally, we note from Fig.~\ref{fit} that for small values of $V_0$ the mean time of collapse is reasonably well fitted to 
\[
\tau_c =718\, V_0^{-2/3},
\]
in units $\bar{\tau}=1$.

\section{Conclusion}

Previous work has demonstrated that vacuum fluctuations at the Planck scale lead to a rapid ``collapse'' of  a congruence of \emph{massless} particles~\cite{carlip1}. This was shown in two-dimensional dilaton gravity, for which the dilaton field was interpreted as the transverse area of the missing dimensions. Although the Raychaudhuri equation for null expansions does not quite make sense in two dimensions, the expansion of this missing area, $\bar\theta$, still obeys a Raychaudhuri-like equation
\begin{equation}
\label{eq:raylike}
\frac{d\bar\theta}{d\lambda}=-\bar\theta^2-16\pi T_L,
\end{equation}
in which the  role of the curvature in the Raychaudhuri equation (\ref{eq:ray}) is taken over by the stress-energy tensor. 

In this work, in contrast, we considered congruences of \emph{massive} particles in two dimensions. In this case, the usual Raychaudhuri equation works perfectly well, and there is no need to bypass the mediation of the spacetime curvature.
As a result, one can directly relate the convergence of the congruence of geodesics to the fluctuations of the curvature tensor, which in turn are caused by the vacuum fluctuations of the stress-energy tensor.

To do so, we considered a simple two-dimensional dilaton model, in which fluctuations in the vacuum energy density of a quantum field induce fluctuations in the curvature of the spacetime. The fluctuations act independently on each rectangular spacetime patch with dimensions $\Delta x=\Delta t=\bar{\tau}$.  As a result, the curvature scalar in each patch inherits a gamma-like probability distribution that has zero mean, is bounded below by a negative value and has an infinite positive tail.  As usual, a positive  curvature scalar focuses the congruence of geodesics, while a negative curvature scalar defocuses. Our numerical results show that, in the end, the focusing always wins, so the congruence always collapses. This agrees with the results of our earlier work~\cite{carlip1}.

It is worth emphasizing that the overall effect of the quantum fluctuations is to focus geodesics, a positive curvature effect. Thus one can think of the quantum fluctuations as effectively generating a small positive {curvature}, at least in the model presented here.

\acknowledgements

S.C. received support from U.S. Department of Energy grant
DE-FG02-91ER40674. R.A.M. and J.P.M.P. acknowledge support from FAPESP grant 2013/09357-9. R.A.M. was partially supported by Conselho Nacional de Desenvolvimento Cient\'{i}fico e Tecnol\'{o}gico under grant 310403/2019-7. J.P.M.P. also acknowledges support from FAPESP Grant No. 2016/07057-6.


\begin{thebibliography}{99}
\bibitem{endnote1}
We also note that the estimates for $\rho_{max}$ and $R_{max}$ in terms of $V_0$  are quite conservative in the sense that the particles themselves always travel inside a thin strip corresponding to $x(t)\ll 1$ (cf Fig.~\ref{traj}), for which the values of $R$ are far smaller than $R_{max}$ as calculated there.
\bibitem{endnote2}
This is exactly the case for distributions with a lower bound (like the gamma distribution or any discrete  distribution with a finite number of possible outcomes) but also correct enough for any reasonable distribution like the Gaussian distribution, for which the probability of drawing a value less than, say, minus ten times the standard deviation is essentially zero.
\end{thebibliography}
\end{document}